\documentclass[twocolumn,preprintnumbers, notitlepage,amsmath,superscriptaddress,10pt,aps,pra]{revtex4-1}

\usepackage{graphicx}
\usepackage{color}
\usepackage{subfigure}
\usepackage{graphicx}
\usepackage{type1cm}
\usepackage{eso-pic}
\usepackage{color}
\usepackage{amsmath}
\usepackage{amssymb}
\usepackage[left]{lineno}
\usepackage{blindtext}
\usepackage{framed}

\begin{document}



\title{Screening nuclear field fluctuations in quantum dots for\\ indistinguishable photon generation}

\date{\today}
\author{R. N. E. Malein}
\thanks{R. N. E. Malein and T. S. Santana contributed equally to this work.}
\author{T. S. Santana}
\thanks{R. N. E. Malein and T. S. Santana contributed equally to this work.}
\author{J. M. Zajac}
\author{A. C. Dada}
\author{E.~M.~Gauger}
\affiliation{Institute of Photonics and Quantum Sciences, SUPA,
Heriot-Watt University, Edinburgh, United Kingdom}
\author{P. M. Petroff}
\affiliation{Materials Department, University of California, Santa Barbara, USA}

\author{J. Y. Lim}
\author{J. D. Song}
\affiliation{Center for Opto-Electronic Convergence Systems, Korea Institute of Science and Technology, Seoul,  Korea}

\author{B. D. Gerardot}
 \email{b.d.gerardot@hw.ac.uk}
\affiliation{Institute of Photonics and Quantum Sciences, SUPA,
Heriot-Watt University, Edinburgh, United Kingdom}


\begin{abstract} 

A semiconductor quantum dot can generate highly coherent and indistinguishable single photons. However, intrinsic semiconductor dephasing mechanisms can reduce the visibility of two-photon interference. For an electron in a quantum dot, a fundamental dephasing process is the hyperfine interaction with the nuclear spin bath. Here we directly probe the consequence of the fluctuating nuclear spins on the elastic and inelastic scattered photon spectra from a resident electron in a single dot. We find the nuclear spin fluctuations lead to detuned Raman scattered photons which are distinguishable from both the elastic and incoherent components of the resonance fluorescence. This significantly reduces two-photon interference visibility. However, we demonstrate successful screening of the nuclear spin noise which enables the generation of coherent single photons that exhibit high visibility two-photon interference.  

\end{abstract}

\maketitle
\twocolumngrid


Indistinguishable single photons are an essential resource for quantum photonic logic gates and networking~\cite{O'Brien_NatPhot_2009}. Among the various approaches to generate identical light quanta, resonance fluorescence (RF) from a semiconductor quantum dot (QD)~\cite{Ates_PRL_2009, Matthiesen_NatComms_2013, He_NatNano_2013, Voliotis_PRB_2014, Proux_PRL_2015} is one of the most promising for practical technological implementation. The RF spectrum is composed of elastic and inelastic scattered light~\cite{Mollow_PR_1969, Nguyen_APL_2011, Matthiesen_PRL_2012, Konthasinghe_PRB_2012}. The elastic or Rayleigh scattered light, first measured in homodyne absorption experiments on a QD~\cite{Hogele_PRL_2004}, has the first order coherence properties of the laser but the second order coherence properties of the emitter~\cite{Loudon}. This light, which can be imprinted with an arbitrary phase or temporal profile~\cite{Matthiesen_NatComms_2013}, is fundamentally indistinguishable. RF can therefore relax the requirement for a perfectly stable, transform-limited optical transition that is difficult to realize in the solid-state. In particular, compared to non-resonant excitation followed by spontaneous emission, RF helps overcome the relatively slow spectral fluctuations caused by charge noise in the QD environment~\cite{Konthasinghe_PRB_2012, Kuhlmann_NatPhys_2013}.

\begin{figure*}
\includegraphics[width=\textwidth]{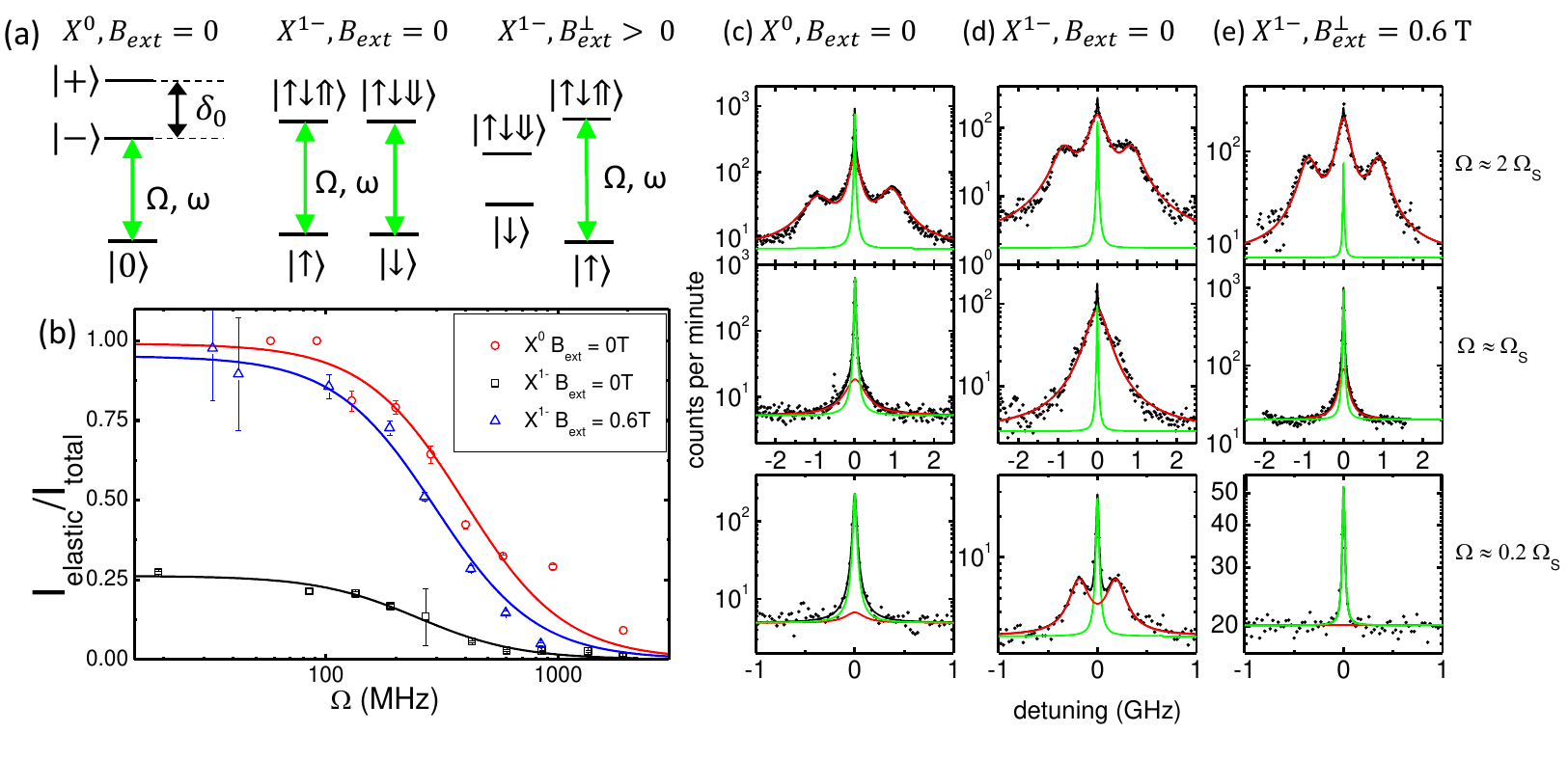}
\caption{\label{fig1}Left: (a) Level diagrams for $X^0$, and $X^{1-}$ at $B_{ext}=0$ T, and $X^{1-}$ at $B_{ext}^{\perp}>0$ T, where $|+\rangle$ and $|-\rangle$ are the symmetric and anti-symmetric neutral exciton eigenstates, respectively; $\delta_0$ is the electron-hole exchange interaction energy, and the green arrows represent the driving fields. (b) Plot of the ratio of elastic scattered photons to the total intensity. The data points are based on the fits to the RF spectra and the curve is a fit using Equation 1 in the main text. Right: high-resolution spectra of (c) $X^0$, and (d) $X^{1-}$ at $B_{ext}^{\perp}=0$ T, and (e) $X^{1-}$ at $B_{ext}^{\perp}=0.6$ T, above ($\Omega > \Omega_S$), near ($\Omega \approx \Omega_S$) and below ($\Omega < \Omega_S$) the saturation Rabi frequency.    The green (red) line is a single (multiple) Lorentzian fit to the elastic (inelastic) component of the RF. The black curve is the total Lorentzian fit. }
\end{figure*}

In addition to the elastic and incoherent components of RF, near resonant excitation can lead to Raman scattering into another ground state. Raman scattering from a QD with a resident electron or hole can have several attractive features. First, below saturation, the coherence of the Raman photons is determined by the ground state dephasing rather than that of the excited state~\cite{Fernandez_PRL_2009, Santori_NJP_2010, He_PRL_2013}. Second, Raman photons are highly tunable: their energy is determined by the detuning of the driving field rather than the fundamental optical transition energy~\cite{Fernandez_PRL_2009, He_PRL_2013, Sweeney_NatPhot_2014}. Finally, due to clean selection rules in a QD, the polarization of the Raman scattered photons can be linked to the spin of the final state~\cite{He_PRL_2013, Sweeney_NatPhot_2014, Yilmaz_PRL_2010}, enabling spin-photon entanglement~\cite{deGreve_Nature_2012, Gao_Nature_2012, Schaibley_PRL_2013} and raising the prospect for quantum networks~\cite{Imamoglu_arXiv_2015}. Typically, spin-flip Raman photons are generated using an in-plane external magnetic field ($B^{\parallel}_{ext}$) to mix the spin states~\cite{Xu_PRL_2007}.    

A III-V QD also contains an intrinsic effective magnetic field, the Overhauser field ($B_{N}$), due to coupling of the QD's constituent atoms' nuclear spins with the electron spin via the contact part of the hyperfine interaction. Incomplete cancellation of the finite random nuclear spin orientations leads to an effective field $B_{N}$ $\approx$ 10 - 30 mT that fluctuates on a $10^{-4}$ s timescale~\cite{Merkulov_PRB_2002, Urbaszek_RMP_2013, Kuhlmann_NatPhys_2013}. The electron spin precession around $B_{N}$ leads to ensemble dephasing on ns timescales~\cite{Merkulov_PRB_2002, Urbaszek_RMP_2013, Braun_PRL_2005}. The fluctuating arbitrary orientation of $B_{N}$ also leads to an almost always present in-plane component $B^{\parallel}_{N}$ which affects the optical properties of a quantum dot. Kuhlmann \textit{et al.} have investigated the role of nuclear spin noise on the neutral ($X^{0}$) and negatively charged ($X^{1-}$) states of a QD~\cite{Kuhlmann_arXiv_2014} while Hansom \textit{et al.} recently exploited $B^{\parallel}_{N}$ to achieve coherent control and two-color coherent population trapping with $X^{1-}$ \cite{Hansom_NatPhys_2014}. Here we report new insights on the effect of $B^{\parallel}_{N}$ on the RF spectrum of a single QD and its consequences on indistinguishable photon generation. We find spin-mixing causes Raman scattered photons clearly distinguishable from the elastic spectrum that reduce the visibility ($\nu$) of two-photon interference (TPI) for $X^{1-}$. We demonstrate successful screening of $B^{\parallel}_{N}$ for both $X^{0}$ and $X^{1-}$ via either an effective magnetic field due to exchange interaction ($\delta_0$) or a perpendicular external magnetic field ($B_{ext}^{\perp}$), respectively. In each case, the spectra exhibit near ideal two-level behaviour and ultra-high visibility TPI is achieved.   

The self-assembled InGaAs QDs studied in this work were embedded in a GaAs Schottky diode for deterministic charge state control. Numerous QDs from two samples at T = 4K were investigated providing consistent results. Sample 1, previously used in Ref.~\onlinecite{Gerardot_APL_2007}, yields $\approx$ 350 kHz from a single QD on a single photon detector at saturation in CW RF. Sample 2 has the identical Schottky diode design but the QDs are positioned at an anti-node of a fifth-order planar cavity on top of a Au layer which functions both as a mirror and Schottky gate. Sample 2 yields $\approx$ 10 times higher count rates due to the planar cavity design~\cite{Ma_JAP_2014}. The RF, obtained using orthogonal linear polarizers in the excitation and collection arms of a confocal microscope~\cite{Matthiesen_NatComms_2013, He_NatNano_2013, Matthiesen_PRL_2012,  Kuhlmann_NatPhys_2013}, is characterized by three techniques: high-resolution (27 MHz) spectroscopy using a Fabry-Perot interferometer (5.5 GHz free spectral range), a Hanbury-Brown and Twiss interferometer to measure $g^2(\tau)$, and an unbalanced Mach-Zender (MZ) interferometer ($\Delta t_{delay} = 49.7$ ns) with polarization control in each arm to measure post-selected, two-photon interference~\cite{Ates_PRL_2009, Proux_PRL_2015}. In the MZ setup, the beamsplitters have nearly perfect 50:50 reflection:transmission. For TPI experiments, a grating (bandwidth $ f/2\pi\sim 1.9$ GHz) is used to spectrally filter the zero-phonon line from the majority of the phonon sideband.  

\begin{figure}
\includegraphics[width=0.5\textwidth]{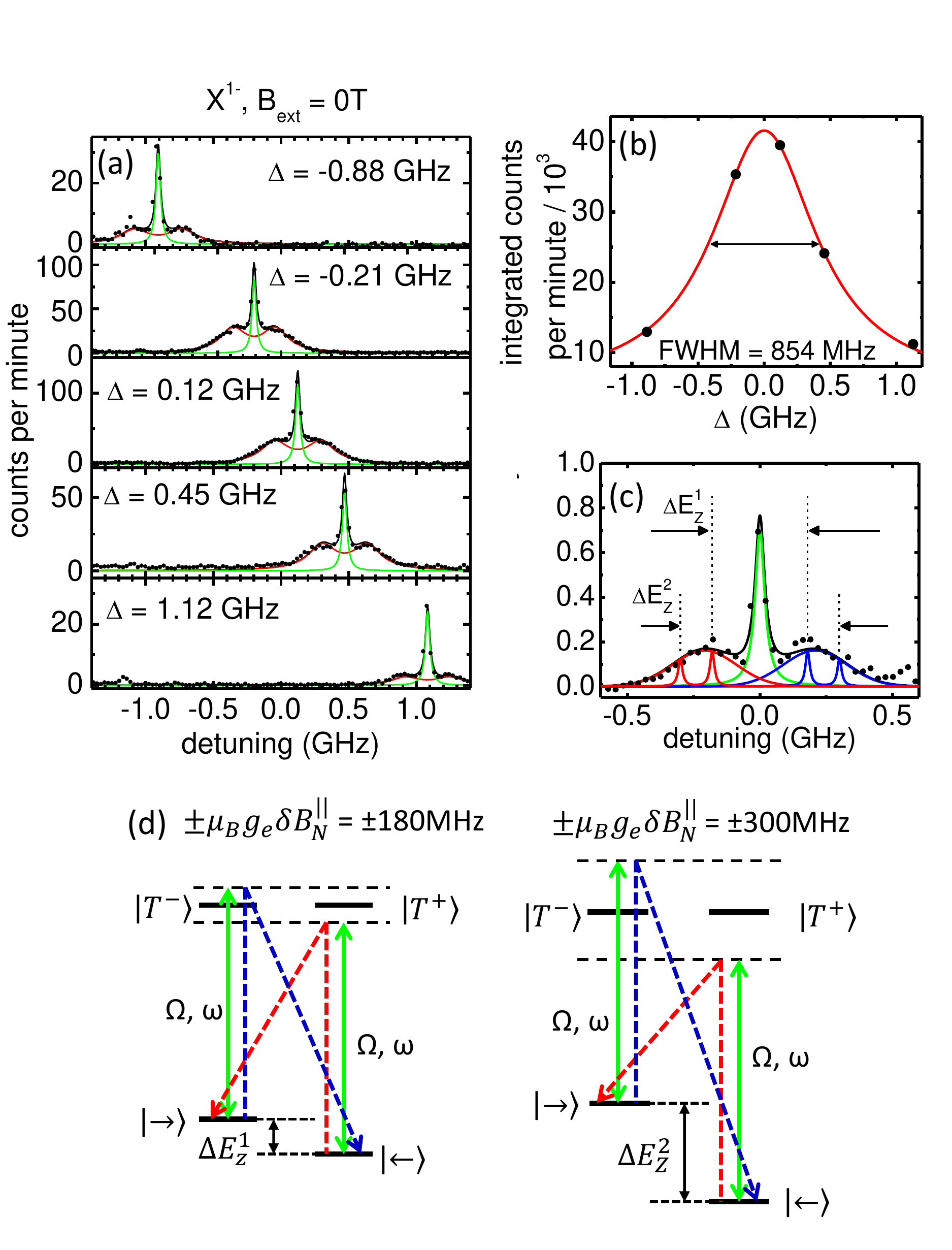}
\caption{\label{fig2} (a) High resolution spectra as a function of laser detuning $(\Delta)$ demonstrate the central and sidebands are dependent on the laser energy. (b) The integrated intensity of the detuned spectra reveal a Lorentzian lineshape (red curve). (c) A simulated spectrum based on the 4-level model including $B_{N}$ (black curve) fit to the experimental data (black points) for  $\Omega =  0.25 \Omega_S$. The green line represents elastically scattered photons.  The red and blue lines represent the red- and blue-detuned Raman photons as shown in the level diagrams for two different $B^{\parallel}_{N}$ values.  (d): Level diagrams for two different values of $B_N^{\parallel}$.  For constant excitation energy, changes in $B_N^{\parallel}$ changes the splitting of the Raman transitions from the laser: fluctuations in the Overhauser field therefore result in broadened Raman sidebands.}
\end{figure}

Figure \ref{fig1} reveals our main spectroscopic result. Here, we compare the resonantly scattered zero-phonon photon spectra from three scenarios: $X^{0}$ and $X^{1-}$ with $B_{ext} = 0$, and $X^{1-}$ with $B_{ext}^{\perp}$ = 0.6 T. The spectra are recorded over a range of Rabi frequencies ($\Omega$), from $\Omega \ll \Omega_S$ where coherent scattering dominates to $\Omega \gg \Omega_S$ where incoherent emission dominates. Here the saturation parameter $\Omega_S$ is given by $\sqrt{1^{}/(T_1 T_2)}$, where $T_1$ is the lifetime and $T_2$ the dephasing time. Fig.~\ref{fig1}~(c) shows that near ideal elastic and inelastic spectra are observed for $X^{0}$, although for $\Omega \gg \Omega_S$ the sidebands of the Mollow triplet are slightly broadened from Mollow's theory due to a modest amount of slow spectral fluctuations (which do not lead to pure dephasing)~\cite{Konthasinghe_PRB_2012}. The central incoherent Lorentzian linewidth is measured as $318 \pm 9.3$ MHz (FWHM), closely matching the Fourier transform limit of the measured $T_1$ value of $550 \pm 40$ ps. Crucially, when the population of the excited state is minimal ($\Omega \ll \Omega_S$), the elastic component dominates the $X^{0}$ spectrum confirming minimal pure dephasing. In Fig.~\ref{fig1}~(b) we plot the ratio of the elastic component to the total spectrum (ER). The solid line is a fit given by~\cite{Mollow_PR_1969, Muller_PRL_2007}:
\begin{equation} 
ER = \dfrac{T_2^*}{2T_1} \dfrac{1}{1 + \Omega^2/\Omega^2_S}
\end{equation} 
where $T_2^*$ is the ensemble dephasing time obtained in time-averaged spectra. By fitting the complete data set we can determine the only unknown parameter, $T_2^*$, accurately and we find $T_2^* = (1.96 \pm 0.08)~T_1 = 1.08 \pm 0.09$ ns for $X^{0}$. 

The spectra for $X^{1-}$ with $B_{ext} = 0$ in Fig.~\ref{fig1}(b) deviate significantly from the near ideal two-level behaviour exhibited by $X^{0}$. All $X^{1-}$ spectra show a reduced elastically scattered component compared to $X^{0}$. Additionally, a doublet surrounding the elastic peak is clearly evident. For $\Omega \ll \Omega_S$, the doublet is separated by $\approx $ 190 MHz and each peak in the doublet has a linewidth of $\approx $ 210 MHz. Fig.~\ref{fig1}(b) shows that the ER saturates at only $\approx 26~\%$ and the fit yields $T_2^* = (0.52 \pm 0.02)~T_1 = 0.40 \pm 0.07$~ns. However, the near ideal two-level behaviour can be recovered with a modest magnetic field in the growth direction, as shown in Fig.~\ref{fig1}(e) for $B^{\parallel}_{ext} = 0.6$T. Based on the fit to the ER, $T_2^* = (1.94 \pm 0.01)~T_1 = 1.51 \pm 0.07$ ns.

\begin{figure}[b]
\includegraphics[width=0.5\textwidth]{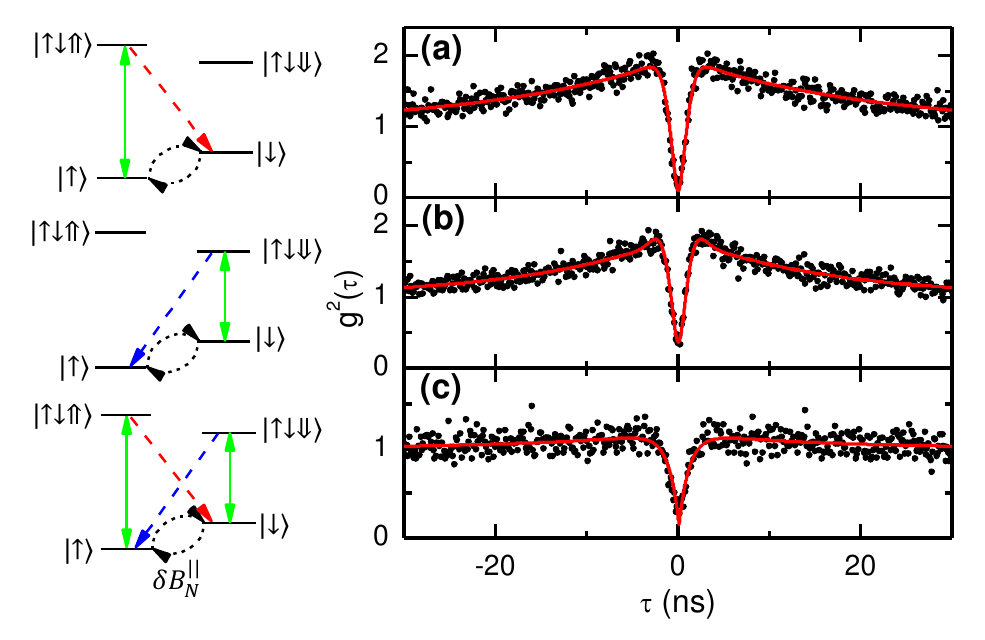}
\caption{\label{fig3} Second-order correlation measurements on the Zeeman-split $X^{1-}$ transitions at $B^{\perp}_{ext} = 0.6$ T and $\Omega = 0.5 \Omega_S$.  Single laser excitation of the (a) blue- and (b) red-shifted lines show bunching around $\tau = 0$ with decay time $\sim 20$ ns.  Simultaneous two-laser excitation (c) shows suppressed bunching, demonstrating that its origin is spin pumping.}
\end{figure}

\begin{figure*}[t]
\centering
\includegraphics[width=0.85\textwidth]{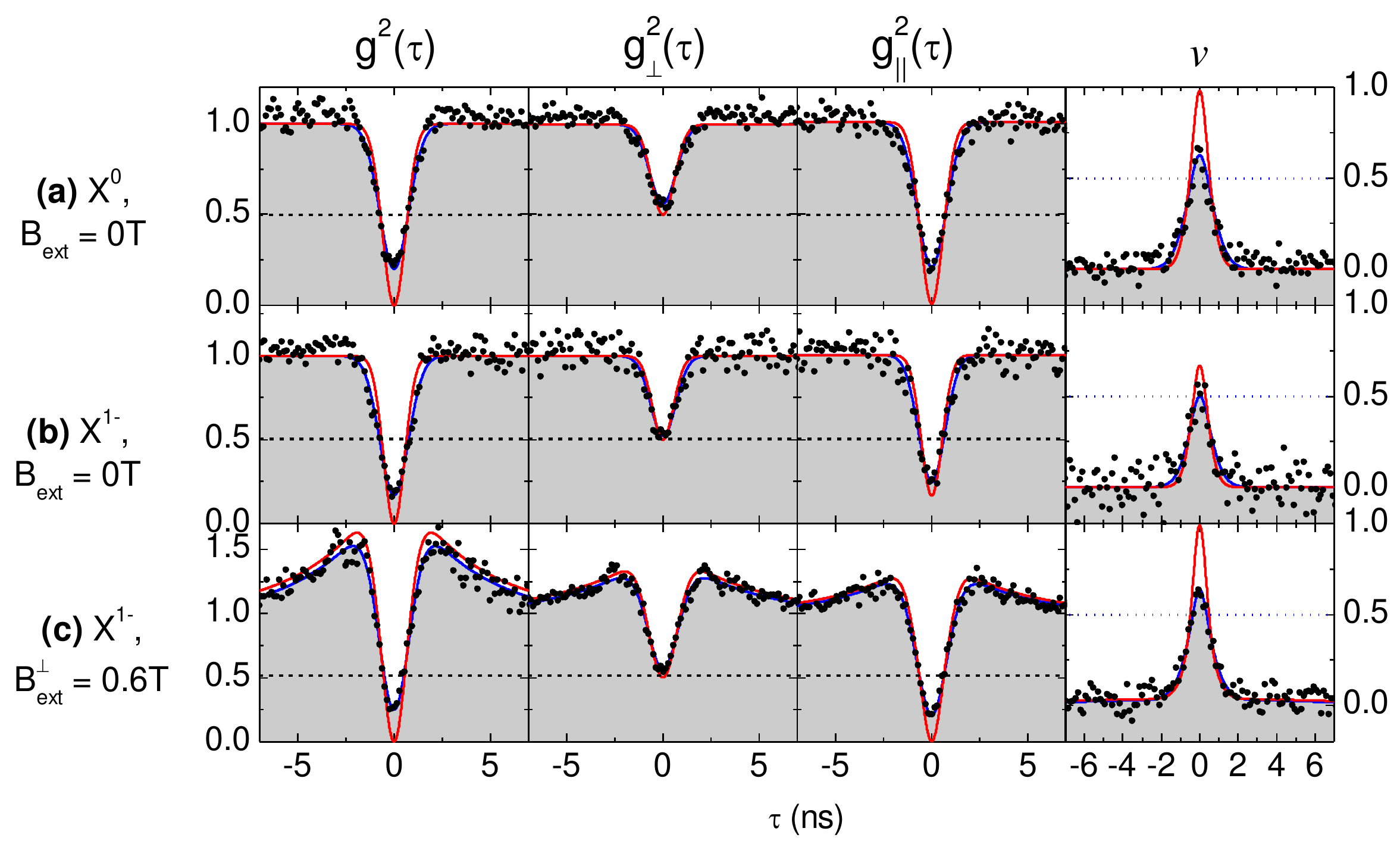}
\caption{\label{fig4} Second-order correlation and two-photon interference measurements on (a) $X^0$, (b) $X^{1-}$ at $B_{ext} = 0$ T, and (c) $X^{1-}$ at $B_{ext}^{\perp} = 0.6$ T.  Blue (red) solid lines are (deconvolved) fits to the data. Near perfect TPI visibility is observed for $X^0$ and $X^{1-}$ with screening of nuclear field fluctuations. For $X^{1-}$ at $B_{ext} = 0$ T, $\nu$ is reduced to $\approx$ 0.67 due to $B_{N}$.}
\end{figure*}

To investigate the origin of the sidebands for $X^{1-}$, we characterize the RF spectrum as a function of laser detuning $\Delta$ for $\Omega \ll \Omega_S$. Here we observe that the entire spectrum follows $\Delta$ (Fig.~\ref{fig2}(a)), and the integrated intensity of the spectrum exhibits a Lorentzian lineshape with a linewidth of $\approx$ 850 MHz (Fig.~\ref{fig2}(b)). This observation eliminates the possible presence of incoherent emission in the spectrum, which is independent of laser detuning. We therefore identify the central peak as the elastic component and the doublet as spin-flip Raman scattering. A schematic of the effect is shown in Fig.~\ref{fig2}(d), in which we depict the four-level system for two values of the $B^{\parallel}_{N}$ component which split the ground-state electron spin states. We assume degenerate excited states as the hyperfine interaction for hole spins is significantly reduced compared to the electron for $B^{\parallel}_{N}$ ~\cite{Loss_PRB_2008, Houel_PRL_2014}. The solid green arrows represent the driving field on resonance with the unperturbed (e.g. $B^{\parallel}_{N}$ = 0) transitions. This driving field determines the elastic scattering spectrum. The red and blue dashed lines represent red- and blue-Raman scattering detuned by $2 \mu_B g_e B^{\parallel}_{N} = \Delta E_Z$. The linewidth of these photons is inversely proportional to the incoherent spin-flip time. The experimental measurement time of a few minutes for each spectrum signifies a time-averaged measurement of $\approx 10^5$ orientations of $B_{N}$, effectively broadening the observed Raman side-bands according to the variance $\delta B^{\parallel}_{N}$. Notably, we also observe QDs which exhibit sidebands with smaller detuning than in Figure 2(a) that can be caused by either reduced $B_{N}$ or $g^{\parallel}_e$. Further details on numerical simulations of the resonantly scattered power spectrum based on the optical Bloch equations of the 4-level system with a linearly polarized driving field are provided in the Supplemental Information.  Numerical simulations to fit the experimental spectra yield a mean energy of $g \mu_B B_{N}$ = $264 \pm 113$ MHz, with a standard deviation $g \mu_B \delta B_{N}$ of $120 \pm 6$ MHz. The black curve in Figure~\ref{fig2}~(c) shows this simulation. Assuming a typical isotropic electron g-factor of -0.6 corresponds to a mean Overhauser field $B_{N}$ = $31 \pm 13$ mT and $\delta B_{N} = 14 \pm 1$ mT.

As the Overhauser field is ubiquitous in III-V QDs, the spectra from both $X^{1-}$ with modest $B_{ext}^{\perp}$ and $X^{0}$ demonstrate successful screening of the nuclear spin fluctuations. In the case of $X^{0}$, the exchange interaction ($\delta_0$) generates a fine-structure splitting of $\approx$ 10 $\mu$eV, much larger than $\mu_B g_e B^{\parallel}_{N} = 0.06$ $\mu$eV. In the case of $B_{ext}^{\perp} \gg B_{N}$ for $X^{1-}$, $B_{ext}^{\perp}$ stabilizes the electron spin which then precesses around  $B_{tot} = B_{ext}^{\perp} + B_{N}$ \cite{Braun_PRL_2005}. This significantly reduces the branching ratio for spin-flip scattering, which can be estimated by $(B_{N})^2 / (2{B_{ext}^{\perp}}^2)$ \cite{Dreiser_PRB_2008}. Hence, the nuclear spin fluctuations are effectively screened and Raman scattered photons are not visible in the spectra of Fig.~\ref{fig1}(e).

As a precursor to TPI experiments, we first investigate the effect of $B_{N}$ on the second-order coherence ($g^2(\tau)$). Fig.~\ref{fig4}(a) shows the auto-correlation from $X^{0}$ with a raw value of $g^2(0) = 0.20 \pm 0.02$, limited by detector jitter. Upon deconvolution of the detector response, $g^2(0) \rightarrow 0$. For $X^{1-}$, while near ideal anti-bunching is again demonstrated, significant bunching around the $\tau$ = 0 dip is observed at $B_{ext}^{\perp} = 0.6$ T (see Fig.~\ref{fig3}~(a)-(b)) due to optical spin pumping caused by electron spin mixing~\cite{Xu_PRL_2007, Hansom_NatPhys_2014, Dreiser_PRB_2008}. The bunching amplitude is determined by the spin pumping fidelity and the decay of the bunching is determined by the spin-relaxation dynamics. A two-color excitation scheme can be used to frustrate the spin pumping mechanism, which unambiguously demonstrates that the nature of the bunching is spin initialization (Fig.~\ref{fig3}(c)).  Although we have observed small bunching amplitudes even at $B_{ext} = 0$ for a few QDs, typically it is negligible for on resonance driving as in Fig.~\ref{fig4} (b).    

The canonical test for photon indistinguishability is two-photon interference, and the visibility is defined as $\nu=[g^2_{\perp}(0)-g^2_{\parallel}(0)]/g^2_{\perp}(0)$, where $g^2_{\perp}(0)$ and $g^2_{\parallel}(0)$ use orthogonal and parallel linear polarizations, respectively, in the two MZ arms. TPI measurements from $X^0$ at $\Omega = 0.5 \Omega_S$ (Fig.~\ref{fig4}(a)) reveal near perfect photon indistinguishability, with a deconvolved (raw) fit value of $\nu = 0.99 \pm 0.02$ ($\nu = 0.58 \pm 0.04$). This impressive result is expected based on the near-ideal high-resolution spectrum, confirming that $\delta_0$ screens $B_{N}$ for $X^{0}$. In contrast, the same measurement on the $X^{1-}$ with $B_{ext} = 0$ T (Fig.~\ref{fig4}(b)) shows reduced $\nu$: a deconvolved (raw) fit value of $\nu = 0.67 \pm 0.04$ ($\nu = 0.50 \pm 0.05$) is measured. Nevertheless, a high degree of indistinguishability is still observed, somewhat surprising considering the spectra in Fig. \ref{fig1}(d). However, in the 'frozen-fluctuation' model, $B_{N}$ is static within the $\Delta t_{delay} = 49.7$ ns, signifying that only one spin-flip Raman transition energy is relevant for TPI. In other words, only narrowband Raman photons will accompany the elastic and modest incoherent components of the RF spectrum over this timescale at $\Omega = 0.5\, \Omega_S$. Finally, applying $B^{\perp}_{ext} = 0.6$ T to screen the ground state electron from the nuclear field fluctuations recovers high visibility TPI with a deconvolved (raw) fit value of $\nu = 0.97 \pm 0.03$ ($\nu = 0.64 \pm 0.06$)(see Fig.~\ref{fig4}(c)). This near perfect indistinguishability meets the expectation based on the spectra in Fig.~\ref{fig1}(e). 

In summary, we have characterized the effect of nuclear field fluctuations on the RF spectrum and photon indistinguishability from $X^0$ and $X^{1-}$ in self-assembled InGaAs QDs.  For the $X^0$, the large $\delta_0$ relative to $B_{N}$ suppresses the effect of nuclear field fluctuations even at $B_{ext} = 0$ T. For a QD with a resident electron in the ground state, the RF spectrum exhibits an inelastic doublet due to spin-flip Raman scattering which reduces the TPI visibility. Application of $B^{\perp}_{ext}$ screens the electron spin from $B^{\parallel}_{N}$, drastically decreasing the rate of Raman scattering and recovering near-ideal indistinguishability.  An interesting prospect would be to investigate the positively charged exciton $X^{1+}$, for which the ground hole spin states can be more robust against nuclear spin noise at $B_{ext} = 0$.

This work was supported by the EPSRC (grant numbers EP/I023186/1, EP/K015338/1, and EP/G03673X/1) and an ERC Starting Grant (number 307392). BDG thanks the Royal Society and EMG thanks the Royal Society of Edinburgh for personal research fellowships.

\newpage

%
\renewcommand{\thefigure}{S\arabic{figure}}
\renewcommand{\theequation}{S\arabic{equation}}
\renewcommand{\thesection}{S\Roman{section}}
\renewcommand{\thetable}{S\arabic{table}}

\newcommand{\ket}[1]{\mathinner{|{#1}\rangle}}
\newcommand{\bra}[1]{\mathinner{\langle{#1}|}}
\newcommand{\braket}[2]{\mathinner{\langle{#1}|{#2}\rangle}}
\newcommand{\ketbra}[2]{\mathinner{|{#1}\rangle\langle{#2}|}}
\newcommand{\ang}{\left(\frac{\theta}{2}\right)}

\newcommand{\Ket}[1]{\left|#1\right>}
\newcommand{\Bra}[1]{\left<#1\right|}
\newcommand{\avg}[1]{\langle #1\rangle}

\newcommand{\SpinDown}{\mathord{\downarrow}}
\newcommand{\SpinUp}{\mathord{\uparrow}}
\newcommand{\TrionUp}{\mathord{\Uparrow}}
\newcommand{\TrionDown}{\mathord{\Downarrow}}

\renewcommand{\thefigure}{S\arabic{figure}}
\renewcommand{\theequation}{S\arabic{equation}}
\renewcommand{\thetable}{S\arabic{table}}

\setcounter{figure}{0}
\setcounter{equation}{0}
\setcounter{section}{0}

\newpage

\onecolumngrid 

\newpage

\onecolumngrid 
{\centering
{\bf \large Supplementary Information: {Screening nuclear field fluctuations in quantum dots for indistinguishable photon generation}}

}

\vspace{20pt}

\makebox[\textwidth][c]{
\begin{minipage}[c]{0.75\textwidth}
{\small { \hspace{6pt}
In this supporting document we provide the following: details on the determination of two-photon interference visibility, modelling of the bunching due observed in second-order correlations due to optical spin pumping of the charged exciton, and  a theoretical model for the resonance fluorescence spectrum of the X$^{1-}$ transition under non-zero nuclear field. This model provides the  fitted spectrum shown in Fig.~2(c) of the main text.
 } }
\end{minipage}
}

\vspace{30pt}

\twocolumngrid

\section{Determination of two-photon interference visibility}
\subsection{Deconvolution}
Determining the indistinguishability based on the experimental data is achieved using the canonical functions for two-photon interference (TPI)~\cite{Legero_APB_2003,Patel_PRL_2008,Proux_PRL_2015}, assuming 50\% reflection and transmission at both beamsplitters:
\begin{align}
g^2(\tau) &= 1 - e^{-|\tau|/\tau_1}\label{eq:g2} \\
g^2_{\perp}(\tau) &=  \frac{1}{4}(g^2(\tau-\Delta \tau) + g^2(\tau+\Delta \tau)) + \frac{1}{2}\, g^2(\tau)\label{eq:cross} \\
g^2_{||}(\tau) &= \frac{1}{4}(g^2(\tau-\Delta \tau) + g^2(\tau+\Delta \tau))(1 - v\, e^{-2|\tau|/\tau_c})\nonumber\\
&+\frac{1}{2} \,g^2(\tau)\label{eq:para}
\end{align}
where $\tau_1$ is the exciton fluorescence lifetime, $\Delta t$ is the fibre delay in the Mach-Zehnder interferometer, $v$ is the visibility and  $\tau_c$ is the photon coherence time. 

These functions are then convolved with the set-up instrument response function (IRF), which is determined to be well approximated by a Gaussian peak with full width at half maximum (FWHM) of 750 ps.  This yields an analytical expression for the convolved function in terms of real physical parameters. We use this convolved function to fit the raw data, and the deconvolved fit can be plotted by using these parameters extracted from the fit in Eqs.~(\ref{eq:g2}-\ref{eq:para}). $\tau_1$ is extracted from a fit to the second order coherence function using $g^2(\tau)$, and is then used as a fixed parameter in the TPI fitting.

\begin{figure*}
\centering
\includegraphics[width=0.9\textwidth]{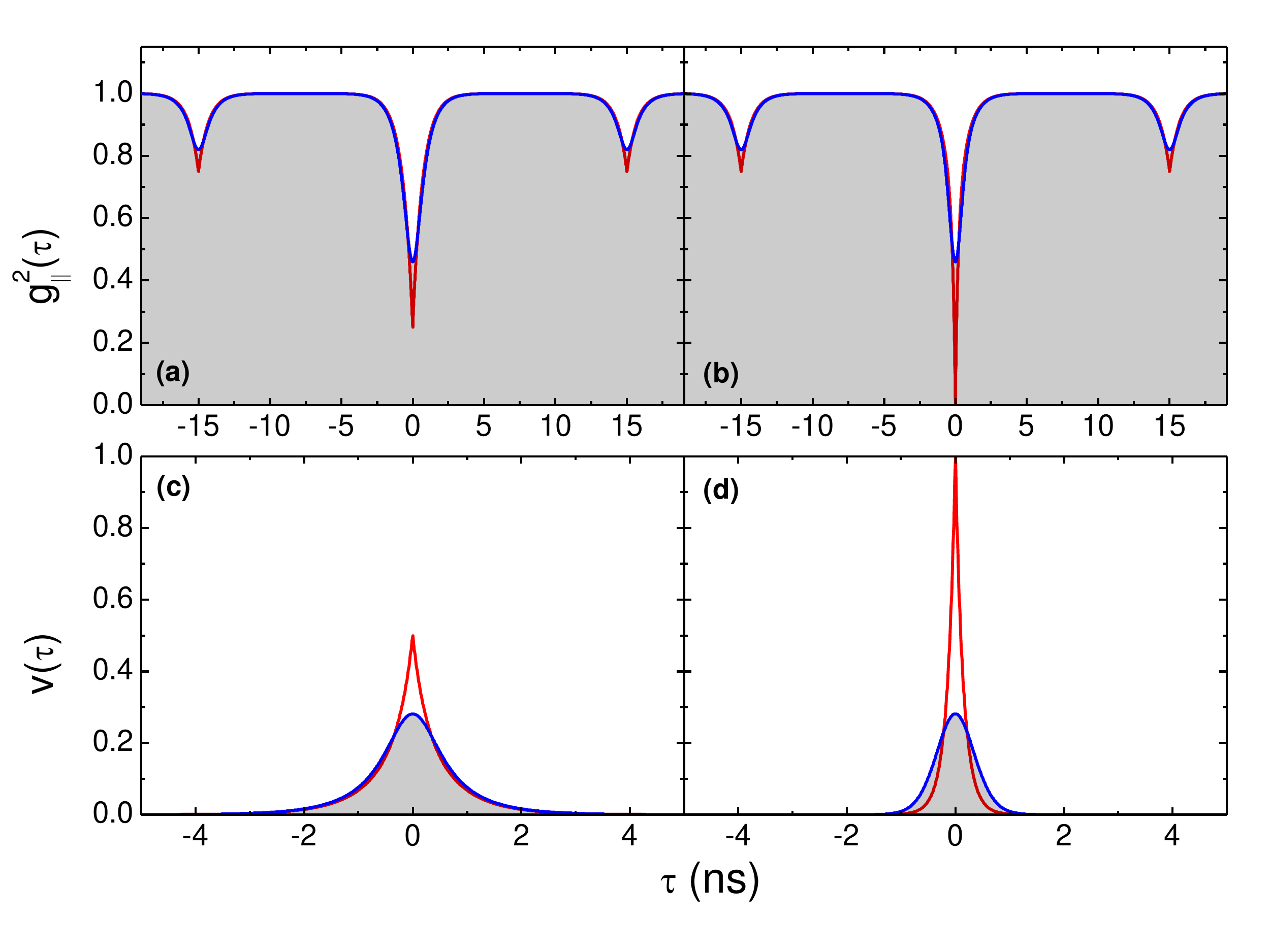}
\caption{\label{fig:S1} Simulated TPI and visibility plots for two different situations showing similar convolved shapes.  Red: plot given by Eqn.~\ref{eq:para}. Blue: plot given by convolving Eqn.~\ref{eq:para} with a Gaussian IRF with FWHM 700 ps. $\tau_1 = 700$ ps, $\Delta \tau = 15$ ns. (a): TPI data with good coherence ($\tau_c = 2\,\tau_1$) but poor visibility ($v=0.6$).  (b): TPI data with perfect indistinguishability ($v=1$) but poor coherence ($\tau_c = 0.5\,\tau_1$).  (c) \& (d): Plots of visibility for the cases in (a) and (b), respectively.}
\end{figure*}

In the convolved TPI data, there is a certain amount of ambiguity between the effects of visibility and coherence time on raw dip depth that determines the photon indistinguishability.  As photons become more distinguishable, $v$ drops from 1 to 0 for perfectly distinguishable photons (e.g. photons with perpendicular polarisations, photons with very different energies), which reduces the depth of the central dip for $g^2_{||}(\tau)$ even after deconvolution.  However, when the interacting photons are  indistinguishable but have very short coherence times, the width of the central dip for $g^2_{||}(\tau)$ from 50\% to 0 will significantly narrow. In this case the experiment is more strongly affected by the system IRF \cite{Proux_PRL_2015}.  
This raises the question
whether a dip depth above zero in the raw data is an effect of photon distinguishability or very short coherence time.  Fig.~\ref{fig:S1} demonstrates this ambiguity.  

In this manuscript, near ideal transform limited linewidths are observed in the high resolution spectra. In the case of $X^{1-}$, we show the non-ideal behaviour is due to Raman scattering at energies detuned from the central transform limited peak in the spectra. We therefore assume this is the origin of the reduced raw (experimental) TPI visibility, rather than reduced coherence or pure dephasing.  

\subsection{Spin pumping}

As observed in Fig.~3, spin pumping in the 4-level system leads to bunching in $g^2(\tau)$ and TPI plots.  This is captured in the fitting process by replacing the $g^2(\tau)$ expression given by Eq.~(\ref{eq:g2}) in Eqs.~(\ref{eq:cross}) and (\ref{eq:para}) with an exponential dip multiplied by an exponential peak with background 1 and a longer decay time $\tau_b$:
\begin{align}
g^2_b(\tau) &= (1 - e^{-|\tau|/\tau_1})(1 + b \,e^{-|\tau|/\tau_b}) \label{eq:bunched}
\end{align}
where $b$ is the amplitude of the bunching. In Fig.~3, this yields $b = 0.85 \pm 0.01$ and $\tau_b = 15.9 \pm 0.3$ ns.  

\section{Resonance fluorescence spectrum of the X$^{1-}$ transition under non-zero nuclear field}

The electron and hole wavefunctions of self-assembled III-V quantum dots (QD) extend over a large number of spin bearing nuclei. The nuclear spins generate an effective magnetic field, called the Overhauser field, that couple with the spins via the hyperfine interaction and can lead to relaxation and dephasing. In the context of the low temperature experiments performed in this manuscript, the ``frozen-fluctuation model''~\cite{merkulov_PRB_2002}, in which the nuclear spins are considered to be stationary for a timescale ($\approx$ 1 $\mu$s) much greater than the optical transition lifetime ($\approx$ 1 ns), is most relevant for self-assembled quantum dots~\cite{kuhlmann_NatPhys_2013,urbaszek_RevModPhy_2013,warburton_Nat_Mat_2013}. In this picture, the electron spin precesses around the hyperfine field on a ns timescale. Here we are interested in changes to the energy levels and selection rules caused by the random Overhauser field orientation. In particular, we develop a theoretical model for predicting the resonance fluorescence spectrum of the photons scattered from the X$^{1-}$ transition of the QD. Our model features four levels, two of those representing the electron spin of the QD's ground state, and the other two the unpaired (heavy) hole spin states of the first excited state.

\begin{figure}[h]
\centering
\includegraphics[width = 0.8\linewidth]{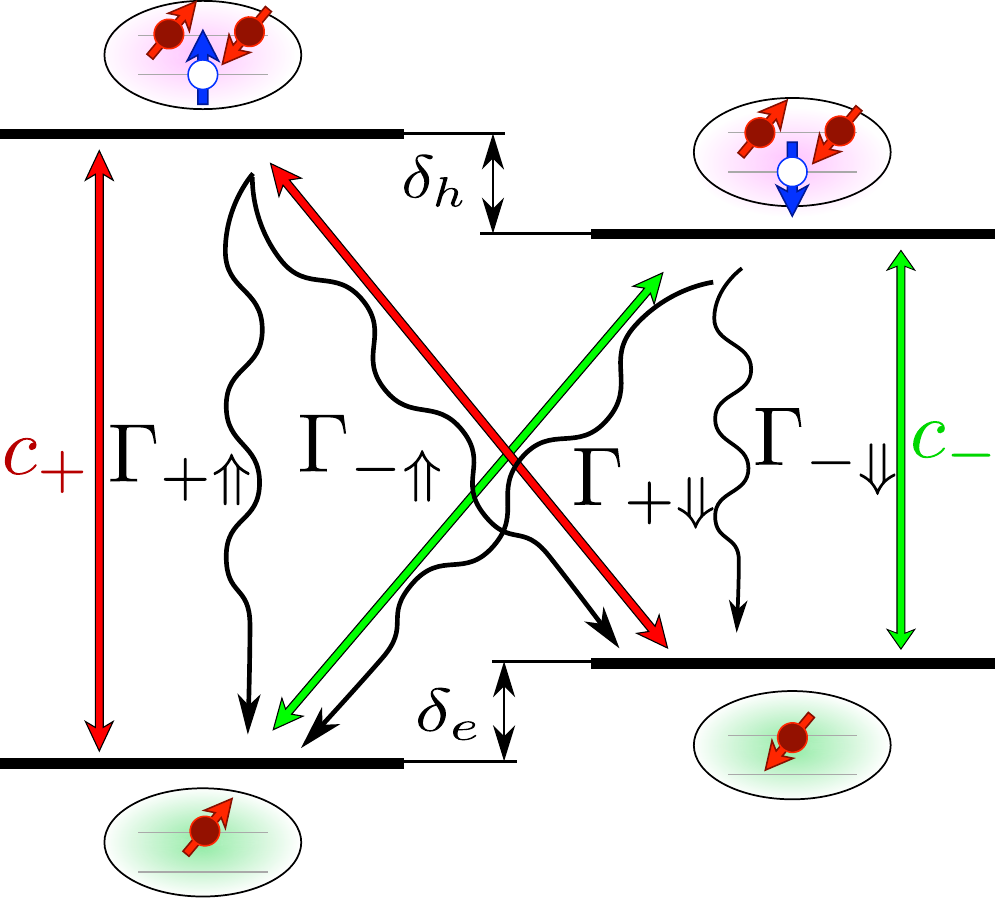}	
\caption{The $X^{1-}$ transition under an Overhauser field with an in-plane component. When the total magnetic field is not completely parallel to the growth ($\hat{z}$) direction the electron spin-mixing alters the selection rules and all four transitions become dipole allowed.}
\label{fig:levels_overhauser}
\end{figure}

We let the electron spin levels be Zeeman split by the total magnetic field, i.e.~the sum of the externally applied field $B_{ext}$ plus the effective Overhauser contribution $B_N$ (in the frozen fluctuation regime).  By contrast, we assume the hole spin only `sees' the external field that is applied along the $\hat{z}$-direction. (Due to the reduced hole Bloch functions at the nuclear positions, the hyperfine interaction coupling coefficient between a hole and nuclear spins is only $\approx$ 10 \% that of the electron spin coupling coefficient~\cite{Fallahi_PRL_2010, Chekhovich_PRL_2010}. Further, the hyperfine interaction for an ideal heavy hole takes on an Ising-like form such that heavy hole spin dephasing is greatly suppressed by an in-plane magnetic field~\cite{Fischer_PRB_2008}). As the basis for our model we adopt the Zeeman eigenstates of both electron and hole (but with respect to different magnetic fields as discussed above), see Fig.~\ref{fig:levels_overhauser} for a schematic depiction of the relevant level structure.

Writing the total magnetic field $\vec{B} = B_{ext} \hat{z} + \vec{B}_{N}$ as
\begin{equation}
\vec{B}=B[\sin\theta\cos\phi\hat{x}+\sin\theta\sin\phi\hat{y}+\cos\theta\hat{z}]
\end{equation}
the electron spin Zeeman ($\vec{\mu} \cdot \vec{B}$) eigenstates are
%
\begin{eqnarray}
\ket{+}&=&\cos\ang\ket{\SpinUp}+e^{i\phi}\sin\ang\ket{\SpinDown} ~,
\\
\ket{-}&=&\sin\ang\ket{\SpinUp}-e^{i\phi}\cos\ang\ket{\SpinDown} ~.
\end{eqnarray}

Denoting the trion states with hole spin (anti)parallel to the $\hat{z}$-direction as $(\Ket{\TrionDown})~\Ket{\TrionUp}$,\cite{note_1} we obtain the following optical dipole transition matrix elements:
\begin{eqnarray}
\bra{\TrionUp}\vec{r}\ket{+}&\propto &\cos\ang\hat{c}_- ~, \label{eq:sel1} \\
\bra{\TrionDown}\vec{r}\ket{+}&\propto &e^{i\phi}\sin\ang\hat{c}_+ ~, \\
\bra{\TrionUp}\vec{r}\ket{-}&\propto &\sin\ang\hat{c}_- ~,\\
\bra{\TrionDown}\vec{r}\ket{-}&\propto &-e^{i\phi}\cos\ang\hat{c}_+ ~, \label{eq:sel4}
\end{eqnarray}
where $\hat{c}_\pm$ refers to the two possible circular polarisation states. Including the dependence of the selection rules on the magnetic field direction and under a rotating wave approximation, the system Hamiltonian is then given by
\begin{widetext}
\begin{eqnarray}
H&=&-\frac{\delta_e}{2}\ketbra{+}{+}+\frac{\delta_e}{2}\ketbra{-}{-}+\left(\Delta-\frac{\delta_h}{2}\right)\ketbra{\TrionDown}{\TrionDown}+\left(\Delta+\frac{\delta_h}{2}\right)\ketbra{\TrionUp}{\TrionUp}\nonumber
\\
&+&\frac{\Omega}{2}\sin\ang(e^{i\phi}\hat{\epsilon}\cdot\hat{c}_+\ketbra{+}{\TrionDown}+H.c.)+\frac{\Omega}{2}\cos\ang(\hat{\epsilon}\cdot\hat{c}_-\ketbra{+}{\TrionUp}+H.c.)\nonumber
\\
&-&\frac{\Omega}{2}\cos\ang(e^{i\phi}\hat{\epsilon}\cdot\hat{c}_+\ketbra{-}{\TrionDown}+H.c.)+\frac{\Omega}{2}\sin\ang(\hat{\epsilon}\cdot\hat{c}_-\ketbra{-}{\TrionUp}+H.c.) ~,
\end{eqnarray}
\end{widetext}
where $H.c.$ denotes the Hermitian conjugate, $\delta_{e}=g_{e}\mu_BB$, $\delta_{h}=g_{h}\mu_BB_{ext}$, $\Omega$ is the optical Rabi frequency and $\Delta$ the laser detuning from the QD transition in the absence of any magnetic field, and $\hat{\epsilon}$ is a unit vector in the direction of the polarization of the driving laser field.

\begin{figure*}[h]
\centering
\includegraphics[width = 0.8\textwidth]{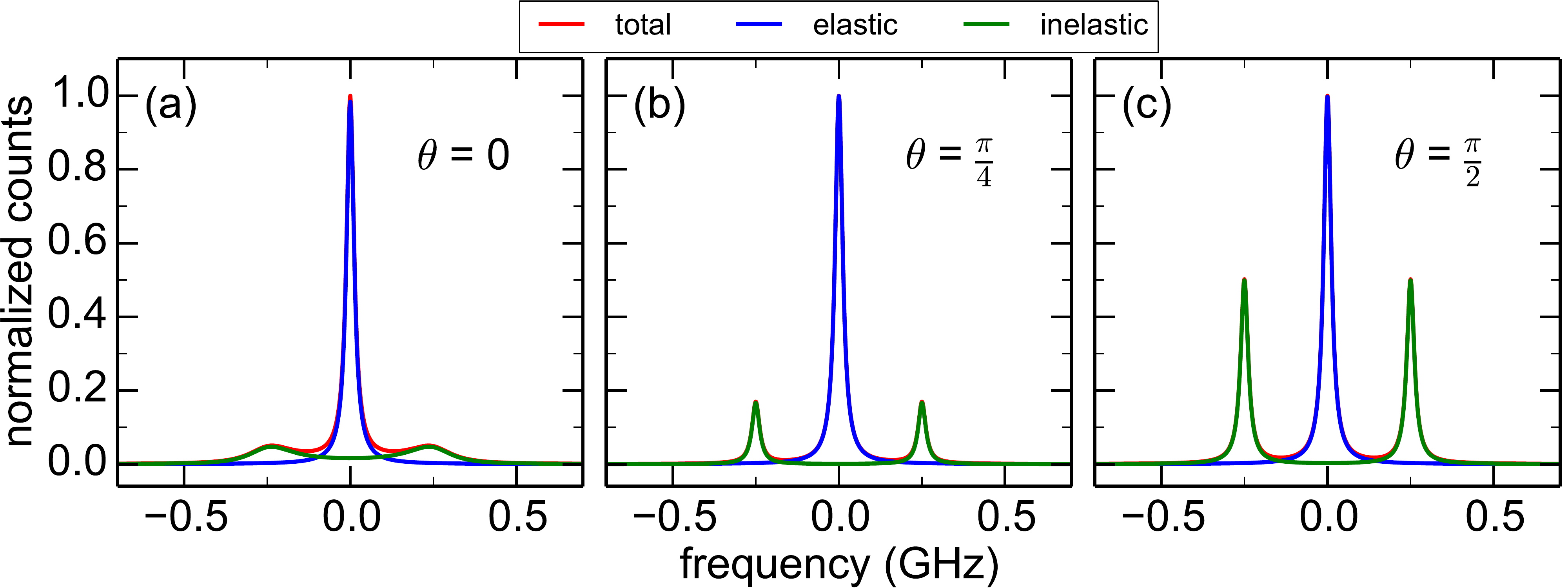}
\caption{Resonance fluorescence spectra with fixed Overhauser fields for $B_{ext}=0$ MHz: (a) Faraday geometry with Rabi frequency $\Omega=0.1\Omega_{sat}$. At sufficiently strong driving (here ten times stronger than in the other two panels of this figure), the central elastic peak is accompanied by two inelastic peaks detuned by $\pm\delta_e/2$ ($B_N=500$ MHz), originating from radiative decay with orthogonal polarisation to that of the excitation photon. (b) $\theta=\pi/4$ and $\Omega=0.01\Omega_{sat}$. The two side peaks at $\pm\delta_e$ ($B_N=250$ MHz) are due to Raman scattered photons (broadened to the interferometer resolution). (c) Voigt geometry with other parameters as in (b): the Raman transitions reach maximum intensity in this configuration.}
\label{fig:OF_single_spec_Gf=0}
\end{figure*}

The selection rules Eqs.~(\ref{eq:sel1}-\ref{eq:sel4}) also imply spontaneous emission decay channels with polar angle $\theta$ dependence as follows
\begin{eqnarray}
\Gamma_{+\Uparrow}&=&\Gamma\cos^2\ang ~, \\
\Gamma_{+\Downarrow}&=&\Gamma\sin^2\ang ~, \\
\Gamma_{-\Uparrow}&=&\Gamma\sin^2\ang ~, \\
\Gamma_{-\Downarrow}&=&\Gamma\cos^2\ang ~,
\end{eqnarray}
where $\Gamma = 1/T_1$ is the inverse lifetime of the QD. These processes are depicted schematically in Fig.~\ref{fig:levels_overhauser}. Further, we define a lowering operator for each circular polarisation
\begin{eqnarray}
\sigma^{(\Downarrow)}_- &=& \left(\sqrt{\Gamma_{+\Downarrow}}\ketbra{+}{\Downarrow}+\sqrt{\Gamma_{-\Downarrow}}\ketbra{-}{\Downarrow}\right)\hat{c}_+ ~,
\\
\sigma^{(\Uparrow)}_- &=& \left(\sqrt{\Gamma_{+\Uparrow}}\ketbra{+}{\Uparrow}+\sqrt{\Gamma_{-\Uparrow}}\ketbra{-}{\Uparrow}\right)\hat{c}_- ~.
\end{eqnarray}

Using these equations, we can obtain the temporal evolution of the density matrix, given by numerically integrating the master equation
\begin{equation}
\frac{d\rho}{dt} = -\frac{i}{\hbar}[H,\rho]+L(\sigma^{(\Uparrow)}_-)\rho + L(\sigma^{(\Downarrow)}_-)\rho ~,
\end{equation}
where $L(\circ) \rho = \circ \rho \circ^\dagger - ( \circ^\dagger \circ  \rho + \rho \circ^\dagger \circ)/2$ is the standard Lindbladian dissipator\cite{footnote_Lind}. 

We proceed to calculate the first-order correlation function $G^{(1)}(\tau) = \avg{\sigma_+(t+\tau)\sigma_-(t)}$, which will give us access to the fluorescence spectrum of the photons scattered by this four-level system. In our case we have (due to the two orthogonal polarisations)
\begin{equation}
G^{(1)}(\tau) = \avg{\sigma_+^{(\Uparrow)}(t+\tau)\sigma_-^{(\Uparrow)}(t)}+\avg{\sigma_+^{(\Downarrow)}(t+\tau)\sigma_-^{(\Downarrow)}(t)} ~.
\end{equation}

The fluorescence spectrum is independent of both the azimuthal angle $\phi$ and the polarisation angle for a linearly polarised driving field. It is calculated by taking the real part of the Fourier transform of the first-order correlation function. For all results presented below, the $\delta$-peaks of elastically scattered photons and the Raman transitions are artificially broadened by making the expected number of photons equal to the area of a Lorentzian curve of width $27.5$ MHz, corresponding to the resolution of the Fabry-Perot interferometer used in the experiments.

The (elastically) scattered Rayleigh peak is present for any orientation of the magnetic field, but its intensity depends on $\theta$. In Faraday geometry ($\theta=0$), there is  no Raman transition and the elastic peak is accompanied by two detuned side peaks at $\pm \delta_e/2$ which possess a width limited by the decay rate of the optical transition [Fig. \ref{fig:OF_single_spec_Gf=0} (a)]. The relative amplitudes of the central to detuned peaks is determined by the Rabi frequency. When the magnetic field is between the Faraday and the Voigt geometry, the relative weight of the two vertical transitions  decreases whilst the two diagonal transitions become allowed. This gives rise to two Raman transitions [Fig. \ref{fig:OF_single_spec_Gf=0} (b)]. These appear as $\delta$-peaks, but have here been broadened by the detector resolution. In the Voigt geometry ($\theta= \pi/2$), all four transitions have equal weight, thus maximising the intensity of the Raman transitions, which now account for $50\%$ of the total number of scattered photons [Fig. \ref{fig:OF_single_spec_Gf=0} (c)].

\begin{figure*}[h!]
\centering
\includegraphics[width = 0.8\linewidth]{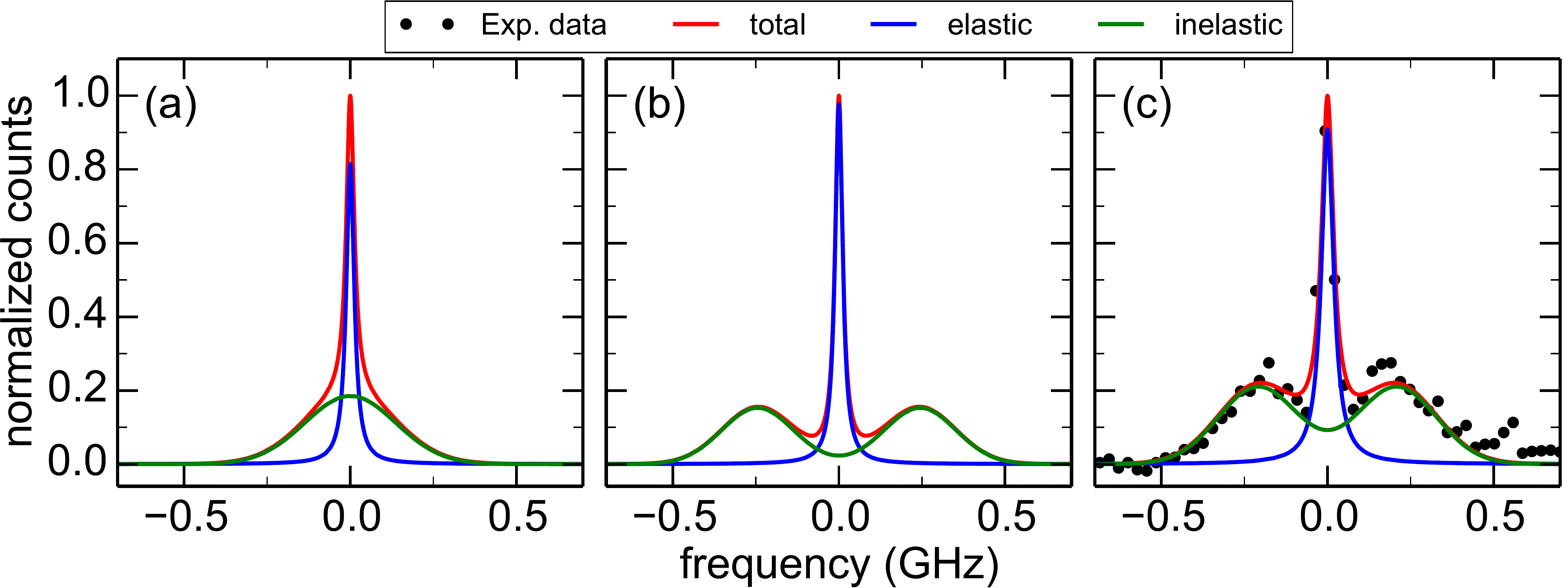}
\caption{Resonance fluorescence spectrum averaged over a fluctuating Overhauser field, using $\Gamma=1.33$ GHz, $\Omega = 0.01\Omega_{sat}$, $B_{ext}=0$ MHz, and an isotropic angular distribution of the Overhauser field throughout. (a) $\avg{B_N}=0$ MHz and $\delta B_N = 200$ MHz: the Raman peak envelope forms a single Gaussian with width proportional to $\delta B_N$. (b) $\avg{B_N} = 300$ MHz and $\delta B_N=100$ MHz: the Raman peak envelopes can be resolved separately either side of the central elastic peak. (c) Our averaged full simulation of the spectrum with $\avg{B_N} = 264$ MHz and $\delta B_N=113$ MHz provides an equally good fit to the experimental data as directly fitting  (i.e.~without underlying calculation of the spectrum) two Gaussian peaks with mean $0.208\pm 0.005$ MHz and standard deviation $0.114\pm 0.006$ MHz plus a Lorentzian peak with mean equal to $0$ MHz and width equal to $27.5$ MHz (FPI resolution). }
\label{fig:OF_noise_spec_Gf=0}
\end{figure*}

To reproduce the long time measurements, it is necessary to average several spectra with different instantaneous nuclear fields $\vec{B}_N$. Considering that the Overhauser field has no preferential direction, we will realise $\avg{ \vec{B}_N}=\vec{0}$ (as required) upon performing uniform averaging over the solid angle. It remains to specify a radial distribution of $B_N$ to represent the magnitudes of the instantaneous Overhauser fields. We consider the case of a normal distribution with generally finite mean $\avg{B_N}$ and variance $\delta B_N$. For $\avg{B_N}=0$ the difference between the energy of the Raman scattered and the Rayleigh photons is typically small, and consequently the average over the Raman peaks yields a single Gaussian envelope of width proportional to $\delta B_N$ [Fig. \ref{fig:OF_noise_spec_Gf=0} (a)]. By contrast a finite mean $\avg{B_N} > 0$ separates the Rayleigh from the Raman peaks, leading to two shifted Gaussian Raman envelopes either side of the central elastic peak [Fig. \ref{fig:OF_noise_spec_Gf=0} (b)]. This latter case gives very close agreement with our measured spectra shown in the main text [see also Fig. \ref{fig:OF_noise_spec_Gf=0} (c)]. 

\begin{thebibliography}{100}

\bibitem{O'Brien_NatPhot_2009} J. L. O'Brien, A. Furusawa, and J. Vu\v{c}kovi\'{c} \textbf{3}, 687 (2009).

\bibitem{Ates_PRL_2009} S. Ates \textit{et al}, Phys. Rev. Lett. \textbf{103}, 167402 (2009).

\bibitem{Matthiesen_NatComms_2013} C. Matthiesen \textit{et al}, Nat. Comms. \textbf{4}, 1600 (2013).

\bibitem{He_NatNano_2013}  Y.-M. He \textit{et al}, Nat. Nanotechno. \textbf{8}, 213 (2013).

\bibitem{Voliotis_PRB_2014} L. Monniell \textit{et al}, Phys. Rev. B \textbf{90}, 041303(R) (2014).

\bibitem{Proux_PRL_2015} R. Proux \textit{et al}, Phys. Rev. Lett. \textbf{114}, 067401 (2015).

\bibitem{Mollow_PR_1969} B. R. Mollow, Phys. Rev. \textbf{188}, 1969 (1969).

\bibitem{Nguyen_APL_2011} H.S. Nguyen \textit{et al}, Appl. Phys. Lett. \textbf{99}, 261904 (2011).

\bibitem{Matthiesen_PRL_2012} C. Matthiesen, A. N. Vamivakus, and M. Atature, Phys. Rev. Lett. \textbf{108}, 093602 (2012).

\bibitem{Konthasinghe_PRB_2012} K. Konthasinghe \textit{et al}, Phys. Rev. B \textbf{85}, 235315 (2012).

\bibitem{Hogele_PRL_2004} A. H\"{o}gele \textit{et al}, Phys. Rev. Lett. \textbf{93}, 217401 (2004).

\bibitem{Loudon} R. Loudon, Oxford University Press, Oxford (2000).

\bibitem{Kuhlmann_NatPhys_2013} A. V. Kuhlmann \textit{et al},  Nat. Phys. \textbf{9}, 570 (2013).

\bibitem{Fernandez_PRL_2009} G. Fernandez \textit{et al}, Phys. Rev. Lett. \textbf{103}, 087406 (2009).

\bibitem{Santori_NJP_2010} C. Santori \textit{et al}, New J. Phys. \textbf{11}, 123009 (2009).

\bibitem{He_PRL_2013}  Y. He \textit{et al}, Phys. Rev. Lett. \textbf{111}, 237403 (2013).

\bibitem{Sweeney_NatPhot_2014} T. M. Sweeney \textit{et al}, Nat. Phot. \textbf{8}, 442 (2014).

\bibitem{Yilmaz_PRL_2010} S. T. Y{\i}lmaz, P. Fallahi, and A. Imamo\u{g}lu, Phys. Rev. Lett. \textbf{105}, 033601 (2010).

\bibitem{deGreve_Nature_2012} K. De Greve \textit{et al}, Nature \textbf{491}, 421 (2012).

\bibitem{Gao_Nature_2012} W. B. Gao, P. Fallahi, J. Miguel-Sanchez, and A. Imamo\u{g}lu, Nature \textbf{491}, 426 (2012).

\bibitem{Schaibley_PRL_2013} J. R. Schaibley \textit{et al}, Phys. Rev. Lett. \textbf{110}, 167401 (2013).

\bibitem{Imamoglu_arXiv_2015} A. Delteil \textit{et al}, arXiv:1507.00465v2 (2015).

\bibitem{Xu_PRL_2007} X. Xu \textit{et al}, Phys. Rev. Lett. \textbf{99}, 097401 (2007).

\bibitem{Merkulov_PRB_2002} I. A. Merkulov, A. L. Efros, and M. Rosen, Phys. Rev. B \textbf{65}, 205309 (2002).

\bibitem{Urbaszek_RMP_2013} B. Urbaszek \textit{et al}, Rev. Mod. Phys. \textbf{85}, 79 (2013).

\bibitem{Braun_PRL_2005} P.-F. Braun \textit{et al}, Phys. Rev. Lett. \textbf{94}, 116601 (2005).

\bibitem{Kuhlmann_arXiv_2014} A. V. Kuhlmann \textit{et al}, Nat Comms. \textbf{6}, 8204 (2015).

\bibitem{Hansom_NatPhys_2014} J. Hansom \textit{et al}, Nat. Phys. \textbf{10}, 725 (2014).

\bibitem{Gerardot_APL_2007} B. D. Gerardot \textit{et al}, Appl. Phys. Lett. \textbf{90}, 221106 (2007).

\bibitem{Ma_JAP_2014} Y. Ma, P. E. Kremer, and B. D. Gerardot, J. Appl. Phys. \textbf{115}, 023106 (2014).

\bibitem{Muller_PRL_2007} A. Muller \textit{et al}, Phys. Rev. Lett. \textbf{99}, 187402 (2007).

\bibitem{Loss_PRB_2008} J. Fischer, W. A. Coish, D. V. Bulaev, and D. Loss, Phys. Rev. B \textbf{78}, 155329 (2008).

\bibitem{Houel_PRL_2014} J. Houel \textit{et al}, Phys. Rev. Lett. \textbf{112}, 107401 (2014).

\bibitem{Dreiser_PRB_2008} J. Dreiser \textit{et al}, Phys. Rev. B \textbf{77}, 075317 (2008).


\end{thebibliography}

\begin{thebibliography}{99}

\bibitem{Legero_APB_2003} T. Legero \emph{et al}, Appl. Phys. B \textbf{77}, 797 (2003)

\bibitem{Patel_PRL_2008} R. B. Patel \emph{et al}, Phys. Rev. Lett. \textbf{100}, 207405 (2008)

\bibitem{Proux_PRL_2015} R. Proux \emph{et al}, Phys. Rev. Lett. \textbf{114}, 067401 (2015)

\bibitem{merkulov_PRB_2002}
I.~Merkulov, A.~L. Efros, and M.~Rosen,
\newblock Phys. Rev. B {\bf 65}, 205309 (2002).

\bibitem{kuhlmann_NatPhys_2013}
A.~V. Kuhlmann {\em et~al.},
\newblock Nature Phys. {\bf 9}, 570 (2013).

\bibitem{urbaszek_RevModPhy_2013}
B.~Urbaszek {\em et~al.},
\newblock Rev. Mod. Phys. {\bf 85}, 79 (2013).

\bibitem{warburton_Nat_Mat_2013}
R.~J. Warburton,
\newblock Nature Mater. {\bf 12}, 483 (2013).

\bibitem{Fallahi_PRL_2010}
P.~Fallahi, S.~Y{\i}lmaz, and A.~Imamo{\u{g}}lu,
\newblock Phys. Rev. Lett. {\bf 105}, 257402 (2010).

\bibitem{Chekhovich_PRL_2010}
E.~Chekhovich {\em et~al.},
\newblock Phys. Rev. Lett. {\bf 104}, 066804 (2010).

\bibitem{Fischer_PRB_2008}
J.~Fischer, W.~Coish, D.~Bulaev, and D.~Loss,
\newblock Phys. Rev. B {\bf 78}, 155329 (2008).

\bibitem{note_1} Note that the paired electron spins are omitted in our notation.

\bibitem{footnote_Lind}
Note that in this particular case, using a single decay dissipator $L(\sigma^{(\Uparrow)}_- + \sigma^{(\Downarrow)}_-)\rho$ gives the same result as our choice of two separate channels.

\end{thebibliography}
\end{document}